\documentstyle[12pt,blois,psfig]{article}
\def\ltsima{$\; \buildrel < \over \sim \;$}
\def\lsim{\lower.5ex\hbox{\ltsima}}
\def\gtsima{$\; \buildrel > \over \sim \;$}
\def\gsim{\lower.5ex\hbox{\gtsima}}
\def\lcdm{$\Lambda$CDM}
\def\lchdm{$\Lambda$CHDM}
\def\hmpc{h^{-1} \, {\rm Mpc}}

\begin{document}

\heading{COLD + HOT DARK MATTER AFTER SUPER-KAMIOKANDE}

\author{Joel R. Primack $^{1}$, Michael A. K. Gross $^{2}$}
{$^{1}$ Physics Department, University of California, Santa Cruz, CA
95064 USA.}
{$^{2}$ Universities Space Research Association, NASA/Goddard Space
Flight Center, Code 931, Greenbelt, MD 20771 USA.}

\begin{figure}[b!]
\footnotesize\centering Invited review at the Xth Rencontres de Blois meeting
\emph{The Birth of Galaxies}, eds. B. Guiderdoni et al, (Gif-sur-Yvette:
Edition Frontieres).
\end{figure}

\begin{bloisabstract}
The recent atmospheric neutrino data from Super-Kamiokande provide
strong evidence of neutrino oscillations and therefore of non-zero
neutrino mass.  These data imply a lower limit on the hot dark matter
(i.e., light neutrino) contribution to the cosmological density
$\Omega_\nu \gsim 0.001$ --- almost as much as that of all the stars in
the universe --- and permit higher $\Omega_\nu$.  The ``standard''
COBE-normalized critical-matter-density (i.e., $\Omega_m=1$) Cold Dark
Matter (CDM) model has too much power on small scales.  But adding to
CDM neutrinos with mass of about 5 eV, corresponding to $\Omega_\nu
\approx 0.2$, results in a much improved fit to data on the nearby
galaxy and cluster distribution.  Indeed, the resulting Cold + Hot Dark
Matter (CHDM) cosmological model is arguably the most successful
$\Omega_m=1$ model for structure formation
\cite{Primack95,Primack96,GawiserSilk98,Gross98}.  However, other recent
data has begun to make the case for $\Omega_m \lsim 0.6$ fairly
convincing.  In light of all this new data, we reconsider whether
cosmology still provides evidence favoring neutrino mass of a few eV in
flat models with cosmological constant $\Omega_\Lambda = 1 - \Omega_m$.
We find that the possible improvement of the low-$\Omega_m$ flat (\lcdm)
cosmological models with the addition of light neutrinos appears to be
rather limited.
\end{bloisabstract} 

\section{Evidence for Neutrino Mass}
Recent articles \cite{GawiserSilk98,Gross98} conclude that, of all the
currently popular cosmological models, the one whose predictions agree
best with the data on the Cosmic Microwave Background (CMB) anisotropies
and the large-scale distribution of galaxies and clusters in the nearby
universe is the $\Omega_m=1$ Cold + Hot Dark Matter (CHDM) model. In
this model, most of the matter (70\% of the total) is cold dark matter,
20\% is hot dark matter, and 10\% is ordinary baryonic matter.  Hot dark
matter is defined as particles that were still moving at nearly the
speed of light at about a year after the Big Bang, when the temperature
was about a keV and gravity first had time to encompass the amount of
matter in a galaxy like the Milky Way; cold dark matter is particles
that were moving sluggishly then.  Few-eV mass neutrinos are the
standard example of hot dark matter.  Three species of neutrinos ---
$\nu_e$, $\nu_\mu$, and $\nu_\tau$ --- are known to exist. The
thermodynamics of the early universe implies that, just as there are
today about 400 CMB photons per cm$^3$ left over from the Big Bang,
there are about 100 per cm$^3$ of each of the three species of light
neutrino (including the corresponding anti-neutrinos).  There are thus
about $4 \times 10^8$ times as many of each species of neutrino as there
are electrons or protons, and as a result a neutrino mass of only 4.7
eV, a mere $10^{-5}$ of the electron's mass, corresponds to 20\% of
critical density in the CHDM model with $h=0.5$.  The relationship
between the total neutrino mass $m(\nu)$ and the fraction $\Omega_\nu$
of critical density that neutrinos contribute is $\Omega_\nu =
m(\nu)/(92 h^2 {\rm eV})$, where $h=0.5-0.8$ is the expansion rate of
the universe (Hubble constant $H_0$) in units of 100 km/s/Mpc.

Direct measurements of neutrino masses have given only upper limits. The 
upper limit on the electron neutrino mass is roughly 10-15 eV; the 
Particle Data Group \cite{Databook} notes that a more precise limit 
cannot be given since unexplained effects have resulted in significantly 
negative measurements of $m(\nu_e)^2$ in recent precise tritium beta 
decay experiments.  There is a (90\% CL) upper limit on an effective 
Majorana neutrino mass of 0.45 eV from the Heidelberg-Moscow $^{76}$Ge 
neutrinoless double beta decay experiment \cite{Klapdor}.  The upper 
limits from accelerator experiments on the masses of the other neutrinos 
are $m(\nu_\mu)< 0.17$ MeV (90\% CL) and $m(\nu_\tau)< 18$ MeV (95\% CL)  
\cite{Databook}, but since stable neutrinos with such large masses would 
certainly ``overclose the universe'' (i.e., prevent it from attaining its 
present age), cosmology implies a much lower upper limit on these 
neutrino masses.  

But there is mounting astrophysical and laboratory data suggesting that 
neutrinos oscillate from one species to another \cite{NuIndustry}, which 
can only happen if they have non-zero mass.  
Of these experiments, the ones that until the new
Super-Kamiokande data were 
regarded as probably most secure were those concerning solar neutrinos.  
But the experimental results that are most relevant to neutrinos as hot 
dark matter are the Liquid Scintillator Neutrino Detector (LSND) 
experiment at Los Alamos and the higher energy atmospheric (cosmic ray) 
neutrino experiments Kamiokande, Super-Kamiokande, MACRO, and Soudan 2. 

Older Kamiokande data \cite{Kam} showed that, for events attributable to 
atmospheric neutrinos with visible energy $E > 1.3$ GeV, the deficit of 
$\nu_\mu$ increases with zenith angle.  The much larger Super-Kamiokande 
detector has confirmed and extended the results of its smaller 
predecessor \cite{SuperK}. These data imply that $\nu_\mu \rightarrow 
\nu_\tau$ oscillations occur with a large mixing angle $\sin^2 2\theta > 
0.82$ and an oscillation length several times the 
height of the atmosphere, which implies that $ 5 \times 10^{-4} < \Delta 
m^2_{\tau \mu} < 6 \times 10^{-3}$ eV$^2$ (90\% CL).  (Neutrino
oscillation experiments measure not the masses, but rather the
difference of the squared masses, of the oscillating species, here
$\Delta m_{\tau \mu}^2 \equiv |m(\nu_\tau)^2 - m(\nu_\mu)^2|$.) 
This in turn 
implies that if
other data requires either $\nu_\mu$ or $\nu_\tau$ to
have large enough mass 
($\gsim 0.5$ eV) to be a hot dark matter particle, then they must be nearly 
equal in mass, i.e., the hot dark matter mass would be shared between 
these two neutrino species.  Both the new Super-Kamiokande atmospheric 
$\nu_e$ data and the lack of a deficit of $\bar\nu_e$ in the CHOOZ 
reactor experiment \cite{chooz} make it quite unlikely that the atmospheric 
neutrino oscillation is $\nu_\mu \rightarrow \nu_e$.  If the oscillation 
were instead to a sterile neutrino, the large mixing angle implies that 
this sterile species would become populated in the early universe and 
lead to too much $^4$He production during the Big Bang Nucleosynthesis 
epoch \cite{shi}.  (Sterile neutrinos are discussed further below.)
It may be possible to verify that $\nu_\mu \rightarrow \nu_\tau$ 
oscillations occur via a long-baseline neutrino oscillation experiment.  
This would look for missing $\nu_\mu$ due to $\nu_\mu \rightarrow 
\nu_\tau$ oscillations with a beam of $\nu_\mu$ from the Japanese KEK 
accelerator directed at the Super-Kamiokande detector, with more powerful 
Fermilab-Soudan and possibly CERN-Gran Sasso long-baseline experiments 
later which could look for $\tau$ appearance.  However, the lower range 
of $\Delta m^2_{\tau \mu}$ favored by the Super-Kamiokande data will make 
such experiments more difficult than was hoped based on the earlier 
Kamiokande data. 

The observation by LSND of events that appear to represent
$\bar\nu_\mu \rightarrow \bar\nu_e$ oscillations followed by $\bar
\nu_e + p \to n + e^+$, $n + p \to D + \gamma$, with coincident
detection of $e^+$ and the 2.2 MeV neutron-capture $\gamma$-ray,
suggests that $\Delta m_{\mu e}^2 > 0$ \cite{lsndan}.  
The independent LSND data \cite{lsndn} suggesting that $\nu_\mu
\rightarrow \nu_e$ oscillations are also occurring is consistent with,
but has less statistical weight than, the LSND signal for $\bar\nu_\mu
\rightarrow \bar\nu_e$ oscillations. Comparison of the latter with
exclusion plots from other experiments allows a range 10 eV$^2 \gsim
\Delta m^2_{\mu e} \gsim 0.2$ eV$^2$.  The lower limit in turn implies
a lower limit $m_\nu \gsim 0.45$ eV, or $\Omega_\nu \gsim 0.02
(0.5/h)^2$.  This implies that the contribution of hot dark matter to
the cosmological density is larger than that of all the visible stars
$\Omega_\ast \approx 0.004$ \cite{Peebles}.  Such an important conclusion
requires independent confirmation.  The KArlsruhe
Rutherford Medium Energy Neutrino (KARMEN) experiment
has added shielding to decrease its background so that it can probe a
similar region of $\Delta m^2_{\mu e}$ and neutrino mixing angle; it
has not seen events attributable to 
$\bar\nu_\mu \rightarrow \bar\nu_e$ oscillations, but the implications 
are not yet clear.
The Booster Neutrino Experiment (BOONE) at Fermilab could
attain greater sensitivity.

The observed deficit of solar electron neutrinos in three different
types of experiments suggests that some of the $\nu_e$ undergo
Mikheyev-Smirnov-Wolfenstein matter-enhanced oscillations $\nu_e
\rightarrow \nu_x$ to another species of neutrino $\nu_x$ with
$\Delta m_{e x}^2 \approx 10^{-5}$ eV$^2$ as they travel through the
sun \cite{solarnu}, 
or possibly ``Just-So'' vacuum oscillations with even smaller
$\Delta m_{e x}^2$ \cite{justso}.
The LSND $\nu_\mu \rightarrow \nu_e$ signal with a much larger $\Delta
m_{e \mu}^2$ is inconsistent with $x=\mu$, and the Super-Kamiokande
atmospheric neutrino oscillation data is inconsistent with $x=\tau$.  Thus
a fourth neutrino species $\nu_s$ is required if all these neutrino
oscillations are actually occurring.  Since the neutral weak boson
$Z^0$ decays only to three species of neutrinos, any additional
neutrino species $\nu_s$ could not couple to the $Z^0$, and is called
``sterile.''  This is perhaps distasteful, although many modern
theories of particle physics beyond the standard model include the
possibility of such sterile neutrinos.  The resulting pattern of
neutrino masses would have $\nu_e$ and $\nu_s$ very light, and
$m(\nu_\mu) \approx m(\nu_\tau) \approx (\Delta m_{e \mu}^2)^{1/2}$,
with the $\nu_\mu$ and $\nu_\tau$ playing the role of the hot dark
matter particles if their masses are high enough \cite{fournu}.  This 
neutrino spectrum might also explain how heavy elements are synthesized 
in core-collapse supernova explosions \cite{caldrev}.
Note that the required solar neutrino mixing angle is very small,
unlike that required to explain the atmospheric $\nu_\mu$ deficit, so a
sterile neutrino species would not be populated in the early universe and 
would not lead to too much $^4$He production.

Of course, if one or more of the indications of neutrino oscillations
are wrong, then a sterile neutrino would not be needed and other
patterns of neutrino masses are possible.  But in any case the
possibility remains of neutrinos having large enough mass to be hot dark
matter.  Assuming that the Super-Kamiokande data on atmospheric
neutrinos are really telling us that $\nu_\mu$ oscillates to $\nu_\tau$,
there are two possibilities regarding neutrino masses:

\noindent {\bf A}) Neutrino masses are hierarchical like all the other
fermion masses, increasing with generation, as in see-saw models.  Then
the Super-Kamiokande $\Delta m^2 \approx 0.002$ implies $m(\nu_\tau)
\approx 0.05$ eV, corresponding to $\Omega_\nu = 0.0015 (m_\nu/0.05 {\rm
eV}) (0.6/h)^2$. This is not big enough to affect galaxy formation
significantly, but it is another puzzling cosmic coincidence that it is
close to the contribution to the density from ordinary matter.

\noindent {\bf B}) The strong mixing between the mu and tau neutrinos
implied by the Super-Kamiokande data suggests that these neutrinos are
also nearly equal in mass, as in the Zee model \cite{Zee} and many
modern models \cite{justso,fournu}.  Then the above $\Omega_\nu$ is
just a lower limit.  An upper limit is given by cosmological structure
formation.  In Cold + Hot Dark Matter (CHDM) models with $\Omega_m=1$,
if $\Omega_\nu$ is greater than about 0.2 the voids are too big
\cite{Ghigna} and there is not enough early structure
(cf. \cite{Primack96} and references therein).

\section{Hot Dark Matter and Structure Formation}
A little hot dark matter can have a dramatic effect on the predicted 
distribution of galaxies \cite{Brodbeck}.  In the early universe the free 
streaming of the fast-moving neutrinos washes out any inhomogeneities in 
their spatial distribution on the scales that will later become galaxies.  
If these neutrinos are a significant fraction of the total mass of the 
universe, then although the density inhomogeneities will be preserved in 
the cold dark matter, their growth rates will be slowed.  As a result, 
the amplitude of the galaxy-scale inhomogeneities today is less with a 
little hot dark matter than if the dark matter is only cold.
(With the tilt $n$ of the primordial spectrum fixed --- which as we
discuss below is not necessarily reasonable --- the fractional reduction
in the power on small scales is $\Delta P/P \approx 8 \Omega_\nu /
\Omega_m$ \cite{Hu98}.)
Since the main problem with $\Omega_m=1$ 
cosmologies containing only cold dark matter is that the amplitude of the 
galaxy-scale inhomogeneities is too large compared to those on larger 
scales, the presence of a little hot dark matter could be just what is 
needed.  And, as was mentioned at the outset, this CHDM model is perhaps 
the best fit to the data on the nearby universe of any cosmological 
model.

But this didn't take into account the new high-z supernova data and
analyses \cite{hizsn} leading to the conclusion that $\Omega_\Lambda -
\Omega_{matter} \approx 0.2$, nor the new high-redshift galaxy data.  As
Somerville, Primack, and Faber \cite{spf} recently found, none of the
$\Omega_m=1$ models with a realistic power spectrum (e.g., CHDM, tilted
CDM, or $\tau$CDM) makes anywhere near enough
bright $z \sim 3$ galaxies.  But we found that \lcdm\ with $\Omega_m
\approx 0.4$ makes about as many high-redshift galaxies as are observed
\cite{spf}.  This value is also suggested if clusters have the same
baryon fraction as the universe as a whole: $\Omega_m \approx \Omega_b /
f_b \approx 0.4$, using for the cosmological density of ordinary matter
$\Omega_b = 0.019 h^{-2}$ \cite{BurlesTytler98} and for the cluster
baryon fraction $f_b = 0.06 h^{-3/2}$ \cite{Evrard97}.  Also a new analysis of
the cluster abundance as a function of redshift is compatible with
$\Omega_m \approx 0.4-1$, with X-ray temperature data favoring $\sim 0.5$
\cite{ecfh,Grossprep} and ENACS and CNOC velocity dispersion data
consistent with
higher $\Omega_m$ \cite{Grossprep}. Thus most likely $\Omega_m$
is $\sim 0.4$ and there is a cosmological constant $\Omega_\Lambda \sim
0.6$.  In the 1984 paper that helped launch CDM \cite{bfpr}, we actually
considered two models in parallel, CDM with $\Omega_m=1$ and \lcdm\ with
$\Omega_m=0.2$ and $\Omega_\Lambda=0.8$, which we thought would bracket
the possibilities.  It looks like an \lcdm\ intermediate between these
extremes may turn out to be the right mix.

\begin{figure}
\centering
\centerline{\psfig{file=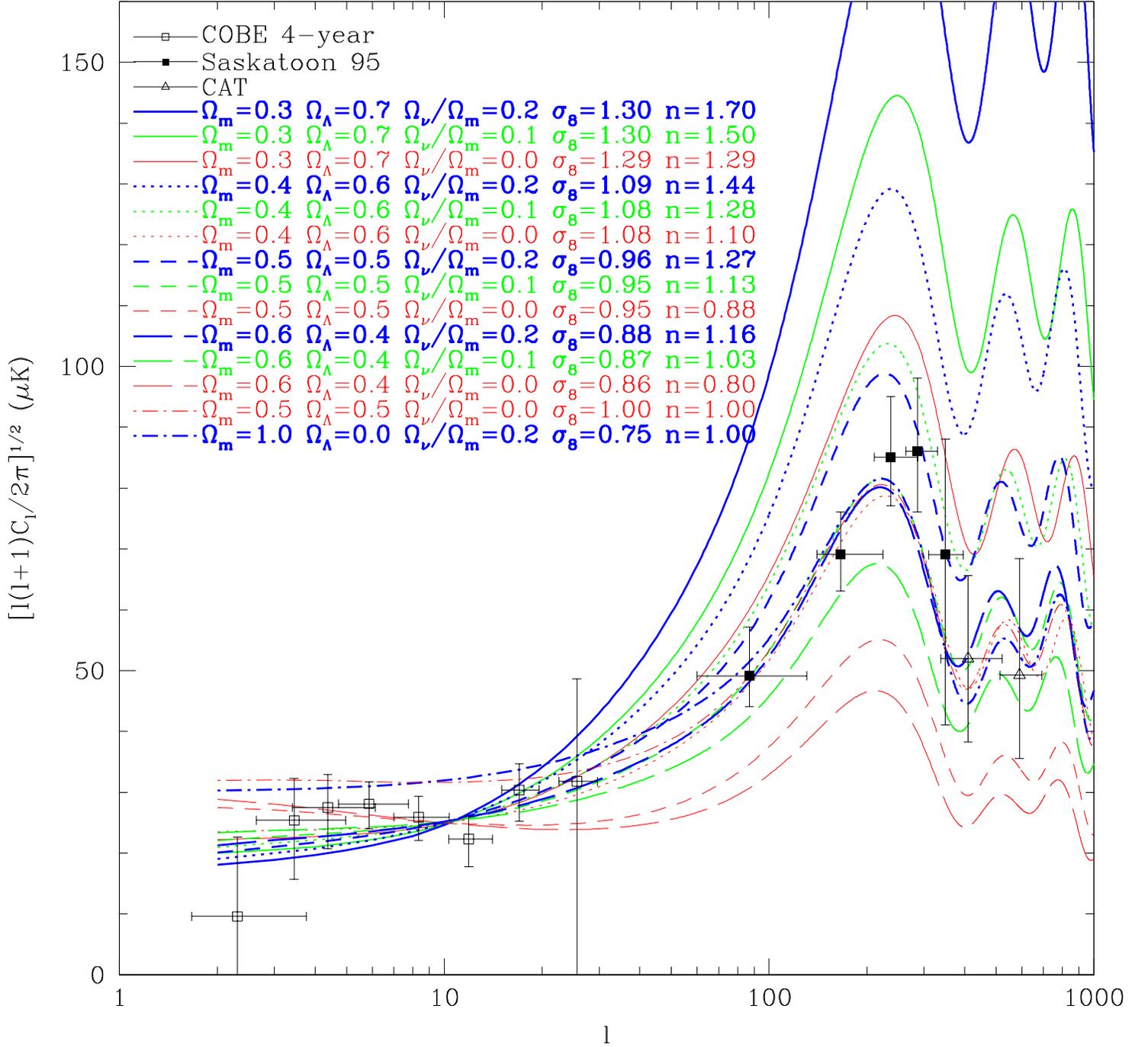}}
\vskip-2pc
\caption{CMB anisotropy power spectrum vs. angular wave 
number for 12 \lchdm\ models with $N_\nu=2$ massive neutrino species 
and Hubble parameter $h=0.6$,
plus the two best-fitting models from Ref. \protect\cite{GawiserSilk98}.
The data plotted are from COBE and two recent small-angle experiments
\protect\cite{sk,cat1,cat2}.
Note that the CHDM model only has 
$N_\nu = 1$ massive neutrino species
because that is what \protect\cite{GawiserSilk98} used.}
\end{figure}

\begin{figure}
\centering
\centerline{\psfig{file=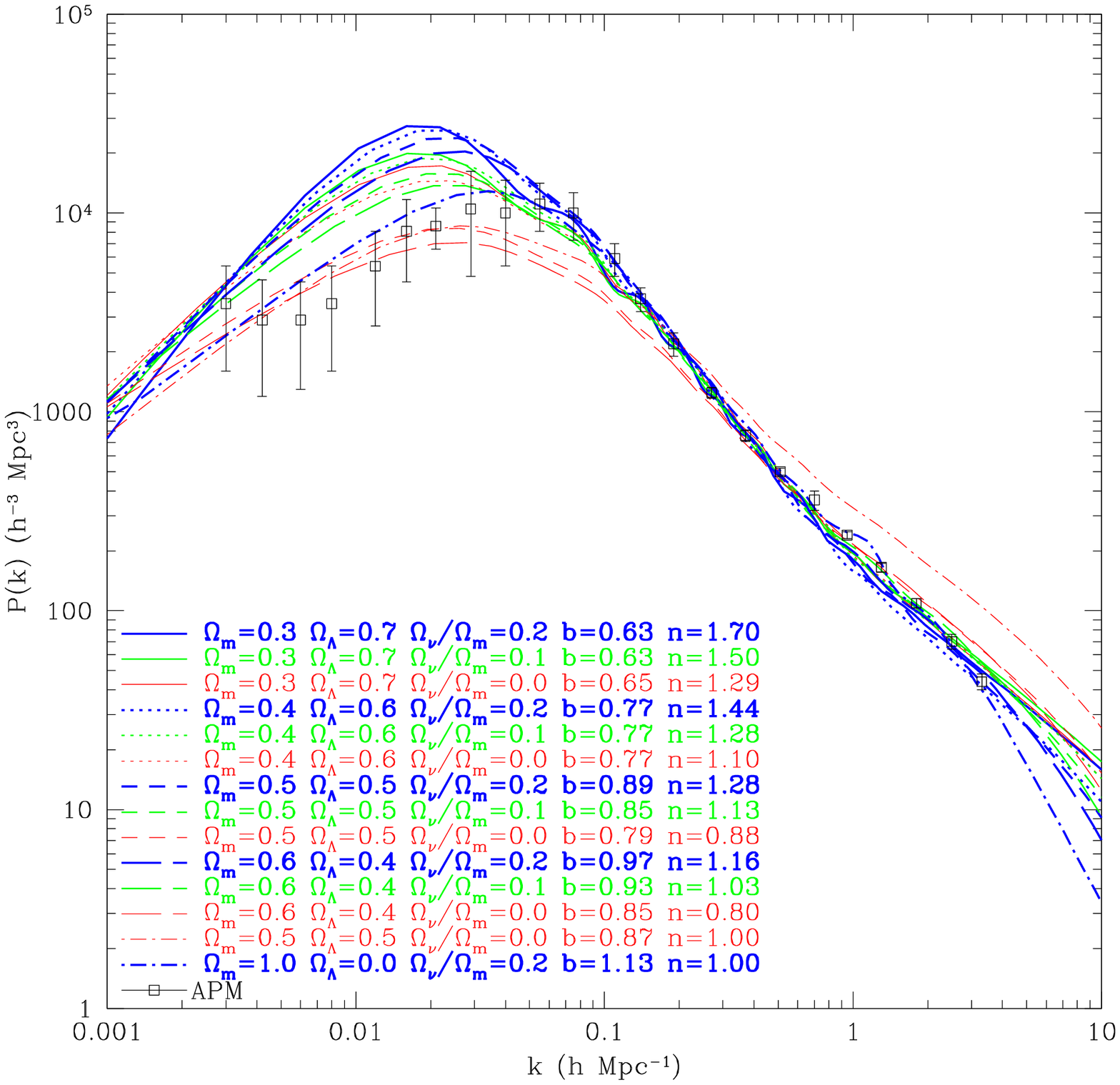}}
\vskip-2pc
\caption{
Nonlinear dark matter power spectrum vs.  wavenumber for the same models
as in Figure 1
Note that we
``nonlinearized'' all the model power spectra \protect\cite{Smith98}, to
allow them all to be compared to the APM data (the ``wiggles'' in the
low-$\Omega_m$ power spectra are an artifact of the nonlinearization
procedure). 
Note also that the last two models differ from the others in that
they are normalized differently and the APM bias is determined over a
smaller range in $k$.  The bias chosen for the $\Lambda$CHDM models is that
which minimizes $\chi^2$ over the entire range of available
APM data.
}
\end{figure}

The success of $\Omega_m=1$ CHDM in fitting the CMB and galaxy
distribution data suggests that low-$\Omega_m$ cosmologies with a
little hot dark matter be investigated in more detail.  We have done
this in order to choose ``best'' models of this \lchdm\ type to
simulate, and we report the results of this investigation here.  We
have used CMBFAST \cite{Seljak} to examine \lchdm\ models with various
$h$, $\Omega_m$, and $\Omega_\nu$, assuming $\Omega_b = 0.019 h^{-2}$,
and we have adjusted the amplitude and tilt $n$ of the primordial
power spectrum $P(k) \propto k^n$ in order to match the COBE amplitude
and the ENACS abundance of clusters \cite{Grossprep}.  (We checked the
CMBFAST calculation of \lchdm\ models against Holtzman's code used in
our earlier investigation of \lchdm\ models \cite{Primack95}.  Our
results are also compatible with those of a recent study
\cite{Valdarnini} in which only $n=1$ models were considered.  But we
find that some \lcdm\ and 
\lchdm\ models require $n>1$, called ``anti-tilt'', and
it is easy to find cosmic inflation models that give $n>1$ ---
cf. \cite{Bono98}.)  The Figures show results for $h=0.6$, where the
simultaneous fit to CMB and APM data is the best of the cases we
considered. Here the neutrino mass is
shared between $N_\nu=2$ equal-mass species --- as explained above, this
is required by the atmospheric neutrino oscillation data if neutrinos
are massive enough to be hot dark matter.  (This results in slightly
more small-scale power compared to $N_\nu=1$ massive
species \cite{Primack95}, but the $N_\nu=1$ curves are very similar to
those shown.)

Of the \lchdm\ models shown, for $\Omega_m=0.5 (0.6)$ the best
simultaneous fits to the small-angle CMB and the APM galaxy power
spectrum data \cite{apm} are obtained for the model with
$\Omega_\nu/\Omega_m=0.1 (0.2)$, and correspondingly $m(\nu_\mu) \approx
m(\nu_\tau) \approx 0.8 (2.0)$ eV for $h=0.6$.
For $\Omega_m \lsim 0.4$, smaller or vanishing neutrino mass appears
to be favored.
Since as mentioned above, the high-$z$ supernovae and other data favor
$\Omega_m \approx 0.4$, we have run a supercomputer simulation of
\lchdm\ with $\Omega_\nu/\Omega_m=0.1$, which is in agreement with the
estimated nonlinear power
spectrum in Figure 2.  Note that the anti-tilt permits some of the \lchdm\
models to give a better fit to the COBE plus small-angle CMB data than the two
$n=1$ models plotted, which are the best-fitting models according to
Ref. \cite{GawiserSilk98}.  Of these, the CHDM model is clearly the best
fit to the APM data.  But the \lcdm\ model has too much power at small
scales ($k \gsim 1\hmpc$), as is well known \cite{KPH} (although recent
work \cite{Colin} suggests that the distribution of {\it dark matter
halos} in the $\Omega_m=0.3$, $h=0.7$ \lcdm\ model may agree well with
the APM data).  On the other hand, the \lchdm\ models may have too
little power on small scales (high-resolution \lchdm\ simulations may be
able to clarify this).  Thus, adding a little hot dark matter to
the low-$\Omega_m$ \lcdm\ models may improve somewhat their simultaneous
fit to the CMB and galaxy data, but the improvement is not nearly as
dramatic as was the case for $\Omega_m=1$.

Let us end with a further note of caution: all \lcdm\ and \lchdm\ models 
that are normalized to COBE and have tilt compatible with the cluster 
abundance are a poor fit to the APM power spectrum near the peak.  The \lchdm\
models all have the peak in their linear power spectrum P(k) higher and at 
lower $k$ than the currently available data (e.g., from APM).
(The \lcdm\ and pure CHDM in the Figures are those 
in \cite{GawiserSilk98}, which are normalized differently.)
Thus the viability of \lcdm\ or \lchdm\ models with a power-law 
primordial fluctuation spectrum (i.e., just tilt $n$) depends on this 
data/analysis being wrong.  The new large-scale surveys 2dF and SDSS will 
be crucial in giving the first really reliable data on this, perhaps as 
early as next year. 

\acknowledgements{JRP acknowledges support from NASA and NSF grants at
UCSC, and thanks Bruno Guiderdoni for the invitation to present this
talk at Blois and Avishai Dekel for hospitality at Hebrew University
where this paper was finished.
MAKG is grateful to the NASA High Performance Computing \&
Communications project for support and the use of Goddard's T3E and
``Beowulf'' supercomputers.}


%
\vfill

\begin{bloisbib}

\bibitem{Primack95} J.R. Primack, J. Holtzman, A.A. Klypin, D.O. Caldwell 
1995, Phys. Rev. Lett. 74, 2160 

\bibitem{Primack96} J.R. Primack 1997, in {\it Critical Dialogues in 
Cosmology}, ed. N. Turok (World Scientific), p. 535; updated in 
astro-ph/9707285

\bibitem{GawiserSilk98} E. Gawiser, J. Silk 1998, Science 280, 1405;
cf. J.R. Primack, Science 280, 1398

\bibitem{Gross98} M.A.K. Gross, R. Somerville, J.R. Primack,
J. Holtzman, and A.A. Klypin 1998, MNRAS, 301, 81 

\bibitem{Databook} C. Caso et al. 1998, Eur. Phys. J. C3, 1

\bibitem{Klapdor} H. Klapdor-Kleingrothaus, hep-ex/9802007

\bibitem{NuIndustry} A comprehensive summary of neutrino data and ongoing 
and proposed experiments is at the ``Neutrino Industry'' web 
site: http://www.hep.anl.gov/NDK/Hypertext/nuindustry.html

\bibitem{Kam} Y. Fukuda et al. 1994, Phys. Lett B 335, 237 

\bibitem{SuperK} Y. Fukuda et al. 1998, Phys. Rev. Lett. 81, 1562

\bibitem{chooz} M. Apollonio et al. 1998, Phys. Lett. B 420, 397 

\bibitem{shi} X. Shi, D.N. Schramm, \& B.D. Fields 1993,
Phys. Rev. D48, 2563; C.Y. Cardall, G.M. Fuller 1996, Phys Rev D 54,
1260; T.I Izotov, T.X. Thuan 1998, ApJ 500, 188

\bibitem{lsndan} C. Athanassopoulos et al. 1996, Phys. Rev. Lett. 77, 
3082; Phys. Rev. C 54, 2685 

\bibitem{lsndn} C. Athanassopoulos et al. 1997, nucl-ex/9706006, Phys. 
Rev. C, in press; 1998, Phys. Rev. Lett. 81, 1774

\bibitem{Peebles} P.J.E. Peebles, {\it Principles of Physical 
Cosmology} (Princeton University Press, Princeton, 1993), eq. 5.150

\bibitem{solarnu} See, e.g., W.C. Haxton 1995, Ann. Rev. Astro. Astroph. 
33, 459; V. Castellani et al. 1997, Phys. Rep. 281, 309 

\bibitem{justso} S.L. Glashow, P.J. Kernan, L.M. Krauss 1998,
hep-ph/9808470; cf. F. Vissani, hep-ph/9708483, and
H. Georgi, S.L. Glashow 1998, hep-ph/9808293

\bibitem{fournu} D.O. Caldwell, R.N. Mohapatra 1993, Phys. Rev. D 48, 
3259; 1994, Phys. Rev. D 50, 3477; J.T. Peltoniemi, J.W.F. Valle 1993, 
Nucl. Phys. B 406, 409; V. Barger, Weiler, T.J., Whisnant, K. 1998, 
Phys. Lett. B 427, 97; S.C. Gibbons et al. 1998, Phys. Lett. 
B 430, 296

\bibitem{caldrev} D.O. Caldwell 1998, in {\it International Workshop on 
Particle Physics and the Early Universe}, ed. L. Roszkowski (World 
Scientific, Singapore); D.O. Caldwell, G.M. Fuller, Y.-Z. 
Qian, in preparation; cf. G.M. Fuller, J.R. Primack, Y.-Z. Qian 1995, 
Phys. Rev. D 52, 1288 

\bibitem{Zee} A. Zee 1980, Phys. Lett. 93B, 389; 1985, Phys. Lett. B161, 
141; cf. A. Smirnov, in 28th Int'l Conf. on HEP, hep-ph/9611465

\bibitem{Ghigna} S. Ghigna et al. 1994, ApJ, 437, L71; S. Ghigna et al. 
1997, ApJ 479, 580

\bibitem{Brodbeck} The effects are illustrated in the video 
accompanying D. Brodbeck et al. 1998, ApJ 495, 1 

\bibitem{Hu98} W. Hu, D.J. Eisenstein, M. Tegmark 1998, 
Phys. Rev. Lett. 80, 5255

\bibitem{hizsn} S. Perlmutter et al. 1998, Nature 391, 51, erratum 392,
311; P.M. Garnavich et al. 1998, ApJ 493, L53; A.G. Riess et al 1998,
astro-ph/9805201, AJ in press; B.P. Schmidt et al. 1998,
astro-ph/9805200, ApJ in press

\bibitem{spf} R.S. Somerville, J.R. Primack, S.M. Faber 1998, 
astro-ph/9806228, MNRAS in press

\bibitem{BurlesTytler98} S. Burles, D. Tytler 1998, ApJ 499, 699; Space
Sci. Rev. 84, 65

\bibitem{Evrard97} A.E. Evrard 1997, MNRAS 292, 289

\bibitem{bfpr} G.R. Blumenthal, S.M. Faber, J.R. Primack, M. Rees 1984, 
Nature 311, 517

\bibitem{ecfh} V.R. Eke, S. Cole, C.S. Frenk, J.P. Henry 1998, 
astro-ph/9802350, submitted to MNRAS

\bibitem{Grossprep} M.A.K. Gross, R.S. Somerville, J.R. Primack, S.
Borgani, M. Girardi 1998, astro-ph/9711035, in {\it Large
Scale Structure: Tracks and Traces}, Proc. of the 12th Potsdam Cosmology
Workshop (15-20 Sept. 1997), eds. V. M\"uller, S. Gottl\"ober,
J.P. M\"ucket, J.  Wambsganss (Singapore: World Scientific, 1998); and
in preparation.  What we actually did in the present paper is to determine the
tilt and $\sigma_8$ simultaneously from COBE \cite{Seljak} plus the ENACS
differential mass function $n(M)$, M. Girardi, S. Borgani, G. Giuricin,
F. Mardirossian, and M. Mezzetti, astro-ph/9804188 v2, ApJ, in press.

\bibitem{Seljak} U. Seljak, M. Zaldarriaga 1996, ApJ 469, 437

\bibitem{Valdarnini} R. Valdarnini, T. Kahniashvili, B. Novosyadlyj 1998, 
A\&A 336, 11 

\bibitem{Bono98} S.A. Bonometto, E. Pierpaoli 1998, New Astron. 3, 391

\bibitem{apm} C. Baugh, G. Efstathiou 1993, MNRAS 265, 145

\bibitem{KPH} A.A. Klypin, J.R. Primack, J. Holtzman 1996, ApJ 466,
1; A. Jenkins et al. 1998, ApJ 499, 20

\bibitem{Colin} P. Colin, A.A. Klypin, A. Kravtsov, A. Khokhlov 1998,
astro-ph/9809202, submitted to ApJ

\bibitem{sk} C.B. Netterfield et al. 1997, ApJ 474, 47

\bibitem{cat1} P.F.S. Scott et al. 1996, ApJ 461, L1

\bibitem{cat2} J.C. Baker 1997, in Proceedings of the Particle Physics and the
Early Universe Conference, www.mrao.com.ac.uk/ppeuc/proceedings

\bibitem{Smith98} C.C. Smith, A.A. Klypin, M.A.K. Gross, J.R. 
Primack, J. Holtzman 1998, MNRAS 297, 910

\end{bloisbib}
\end{document}